# Dual-layer large-numerical-aperture metalenses with rotational zoom capability*

LEI Tingfeng, SHU Guanqing, CHEN Xiaodong*

School of Physics, Sun Yat-sen University, Guangzhou 510275, China

**Abstract**

Metalenses have emerged as a versatile platform for wavefront engineering, enabling compact and planar optical systems with unprecedented functional density. However, most reported metalenses rely on single-layer architectures that inherently support only a fixed focal length, which limits their application in tunable imaging and adaptive photonic systems. Dual-layer Moiré-type metalenses offer an alternative pathway to achieving zoom functionality by modulating the effective phase profile through relative rotation rather than axial translation. Despite this advantage, existing designs predominantly employ a paraxial parabolic phase approximation based on the truncated expansion of the ideal lens phase. Such approximations become inaccurate under large-numerical-aperture (*NA*) conditions, where neglected higher-order phase components introduce accumulated spherical aberration and degraded focusing performance. Here, we present a non-paraxial design strategy based on the full ideal aberration-free phase function for a dual-layer rotational zoom metalens. By adopting the exact lens phase expression rather than the conventional parabolic approximation, this approach preserves the higher-order phase contributions critical to large-*NA* operation. Zooming is achieved by engineering complementary phase distributions on two metasurface layers and dynamically modulating their combined phase via controlled rotation. To realize the target phase profiles, a dielectric metasurface operating at 10 GHz is designed using ceramic cylindrical resonators arranged in a triangular lattice. A systematic parametric optimization of lattice period, cylinder diameter, and height ensures a complete 2π phase coverage while maintaining a transmission above 85%. The resulting phase-geometry mapping establishes a reliable unit-cell library for device implementation. Leveraging this library, the dual-layer metalens is designed and its performance evaluated through finite-difference time-domain simulations. A prototype with a 300 mm aperture is fabricated and experimentally characterized using three-dimensional near-field scanning measurements. Both simulations and experiments confirm that continuous focal-length tuning is achieved as the rotation angle varies from –60° to 60°, corresponding to a focal-length range of 100 to 300 mm and a zoom ratio of 3∶1. The device achieves a maximum *NA* of 0.83, and the measured focal spots remain nearly diffraction-limited. The experimentally extracted focal-length variation agrees closely with theoretical predictions and simulations. These results validate the non-paraxial design model and demonstrate the viability of large-*NA* zoom operation in a dual-layer metasurface. This work offers a promising route for lightweight variable-focus optics and holds significant potential for next-generation tunable imaging systems and compact optical platforms.

# 1 Introduction

Metasurfaces are two-dimensional artificial materials composed of subwavelength structural units. They have attracted extensive research in the fields of optical and electromagnetic wave manipulation in recent years[1,2]. By precisely designing structural units and geometric arrangements at the subwavelength scale, metasurfaces enable flexible control over light wave properties such as amplitude, phase, and polarization. They are widely applied in wavefront shaping[3,4], holographic displays[5-7], and polarization control[8,9]. These applications demonstrate functional density and integration potential that surpass traditional refractive optical elements. As an important branch of the metasurface family, metalenses achieve the focusing and imaging functions of traditional lenses through high-degree-of-freedom phase control. This provides a novel pathway for realizing compact, planar optical systems[10-13].

However, most traditional metalenses are static optical elements with fixed focal lengths and fixed numerical apertures (*NA*). Their lack of tunability limits their application in systems such as zoom optics and beam steering. To overcome the design limitations of single-layer structures, researchers have proposed bilayer metasurface structures[14-19]. By introducing a second interface with independent phase control capabilities, these structures achieve higher design freedom and enhance wavefront manipulation and light field shaping. Based on this, bilayer rotational zoom metalenses based on the Moiré effect (interference fringe effect) have received widespread attention due to their simple structure, absence of mechanical displacement, and ability to achieve continuous focal length adjustment[20-37]. Such metalenses typically consist of two metasurface layers with specific phase distributions. Rotating one layer around the optical axis changes the superimposed phase distribution of the two layers, thereby controlling the equivalent focal length of the system. Bernet et al.[20,21] proposed the theoretical model of Moiré lenses and experimentally verified their tunable focal length characteristics using diffractive optical elements. Subsequently, this concept was introduced into metasurface systems, achieving significant progress in microwave[22], infrared[23], visible light[25,26,29,36], terahertz[28,34,37], and ultrasonic[35] bands. For instance, Iwami et al.[23] reported a novel ultra-thin planar lens based on a rotationally tunable Moiré metalens. It achieved continuous adjustment from negative to positive focal lengths in the 900 nm near-infrared band, featuring both a large aperture and a wide tuning range. Luo et al.[25] further applied this concept to fluorescence microscopy, achieving high-contrast multi-plane imaging in the visible light band. Additionally, explorations in polarization multiplexing[30,33] and topological charge control[31,32] have continued to expand the functionality of Moiré metalenses. These efforts have enabled various novel optical applications, including dual vortex beam generation[31] and augmented reality displays[36].

Although existing studies have verified the feasibility of Moiré metalenses in focal length tuning and wavefront reconstruction, their performance remains constrained by several key factors. Traditional designs are typically based on a parabolic phase formula model. This phase formula is derived from the paraxial approximation expansion of the ideal lens phase function. It offers good approximation accuracy under paraxial conditions with small angles and low numerical apertures. For high-*NA* systems, ignoring high-order phase terms under large numerical aperture conditions causes the phase in non-paraxial regions to deviate from the ideal aberration-free lens phase. This deviation leads to significant aberrations[38]. The limitations of this theoretical approximation not only reduce focusing efficiency and imaging quality but also restrict the application of rotational zoom lenses under high-*NA* conditions. Therefore, it is necessary to extend the existing theoretical framework of rotational zoom Moiré metalenses. Developing a phase model suitable for large numerical aperture conditions is of great significance for advancing high-performance zoom optical systems.

To address these issues, this paper proposes a design method for bilayer rotational zoom metalenses based on the phase function of an ideal aberration-free lens. We designed verification samples operating at a frequency of 10 GHz. Unlike traditional design approaches that rely on parabolic approximation, this method starts directly from the complete phase function of an ideal lens. It derives the phase expression for the bilayer structure without introducing paraxial approximation assumptions. This theoretically reduces high-order phase errors introduced by paraxial approximation. Both numerical simulations and experimental results indicate that this zoom bilayer metasurface achieves a zoom range of approximately 3:1. The trend of focal length change is consistent with theoretical predictions, and the maximum *NA* reaches 0.83. Starting from the phase function of an ideal aberration-free lens, this study corrects the phase design model based on parabolic approximation in traditional rotational zoom Moiré metalenses. Consequently, it realizes bilayer rotational zoom metalenses under high-*NA* conditions, providing new theoretical and design references for compact planar zoom optical systems.

## 2 Theoretical Model and Design Method

The bilayer rotational zoom metalens consists of two metasurface layers, $\mathrm{MS}_1$ and $\mathrm{MS}_2$, with specific phase distributions, as shown in Figure 1(a). Rotating the second metasurface layer $\mathrm{MS}_2$ around the optical axis adjusts the equivalent focal length of the system, thereby achieving zoom functionality. When designing the phase functions of the bilayer metasurface, a fixed design rotation angle $\alpha_0$ is first set. It is specified that when the second layer rotates by this angle, the system focal length changes from $f_1$ to $f_2$. Here, $\alpha_0$ is a constant during the design phase, used to determine the phase distribution relationship between the two metasurface layers. Let the phase functions of the two metasurface layers $\mathrm{MS}_1$ and $\mathrm{MS}_2$ be $\varphi_1(r,\theta)$ and $\varphi_2(r,\theta)$, respectively. The combined phase of the system before and after rotating the second layer by the angle $\alpha_0$ is given by

$$\varphi_{\text{total}}(r,\theta) = \varphi_{f_1}(r), \tag{1}$$

$$\varphi_{\text{total}}(r,\theta+\alpha_0) = \varphi_{f_2}(r), \tag{2}$$

Here, $\varphi_{f_1}(r)$ and $\varphi_{f_2}(r)$ describe the phase distributions of lenses with focal lengths $f_1$ and $f_2$, respectively, before and after system rotation. Subtracting the two equations yields

$$\varphi_2(r,\theta+\alpha_0) - \varphi_2(r,\theta) = \varphi_{f_2}(r) - \varphi_{f_1}(r). \tag{3}$$

To facilitate the separation of angular variables, let $\varphi_2(r,\theta) = g(r)\cdot\theta + f(r)$, where $g(r)\cdot\theta$ represents the angle-dependent term and $f(r)$ denotes an arbitrary circularly symmetric function. Substitution into the above equation yields

$$g(r)\cdot\alpha_0 = \varphi_{f_2}(r) - \varphi_{f_1}(r). \tag{4}$$

Since radially symmetric functions do not affect the lens focusing results, we can set $f(r) = 0$. Thus, the phase function of the second-layer metasurface can be expressed as

$$\varphi_2(r,\theta) = \frac{\theta}{\alpha_0}[\varphi_{f_2}(r) - \varphi_{f_1}(r)]. \tag{5}$$

It is important to note that, in polar coordinates, the phase function exhibits a phase jump at $\theta = \pi$ that is not an integer multiple of $2\pi$, thereby violating the $2\pi$ periodicity of the optical phase. To eliminate this discontinuity, we applied a rounding correction (using the round function) to the angle-dependent term $g(r)$, ensuring continuous phase variation along the angular direction[20]. The resulting phase function expressions for the two-layer metasurface are as follows:

$$\varphi_1(r,\theta) = \varphi_{f_1}(r) - \varphi_2(r,\theta), \tag{6}$$

$$\varphi_2(r,\theta) = \text{round}[\frac{\varphi_{f_2}(r) - \varphi_{f_1}(r)}{\alpha_0}] \cdot \theta. \tag{7}$$

Next, we need to determine the phase function expressions for $\varphi_{f_1}(r)$ and $\varphi_{f_2}(r)$. The phase function of an ideal focusing lens is

$$\varphi_{\text{ideal}}(r) = -\frac{2\pi}{\lambda_0}(\sqrt{r^2+f^2} - f). \tag{8}$$

Expand using the paraxial approximation ($r \ll f$):

$$\sqrt{r^2+f^2} \approx f + \frac{r^2}{2f} - \frac{r^4}{8f^3} + \cdots,$$

After neglecting higher-order terms, the parabolic approximation of the lens phase distribution is obtained:

$$\varphi_{\text{parabolic}}(r) = -\frac{\pi r^2}{\lambda_0 f}. \tag{9}$$

The parabolic approximation of the lens phase function applies to systems with small *NA*. However, when $r$ approaches $f$, specifically under large *NA* conditions, the influence of high-order phase terms on wavefront propagation becomes non-negligible. Off-axis rays fail to focus conjointly with on-axis rays, thereby inducing aberrations and causing significant degradation in focusing performance. To achieve superior focusing effects under large *NA* conditions, we adopt the phase function of the ideal aberration-free lens from Eq. (8) as the phase expressions for $\varphi_{f_1}(r)$ and $\varphi_{f_2}(r)$. This approach theoretically eliminates phase errors introduced by the paraxial approximation, thus enhancing the applicability of the dual-layer zoom metalens model under large *NA* conditions. The final phase distribution of the dual-layer metasurface is shown in Fig. 1(b). As the rotation angle $\alpha$ increases from 0 to 60°, the number of equiphase circles in the synthesized phase decreases, corresponding to an increase in focal length. This observation validates the zooming functionality of the system.

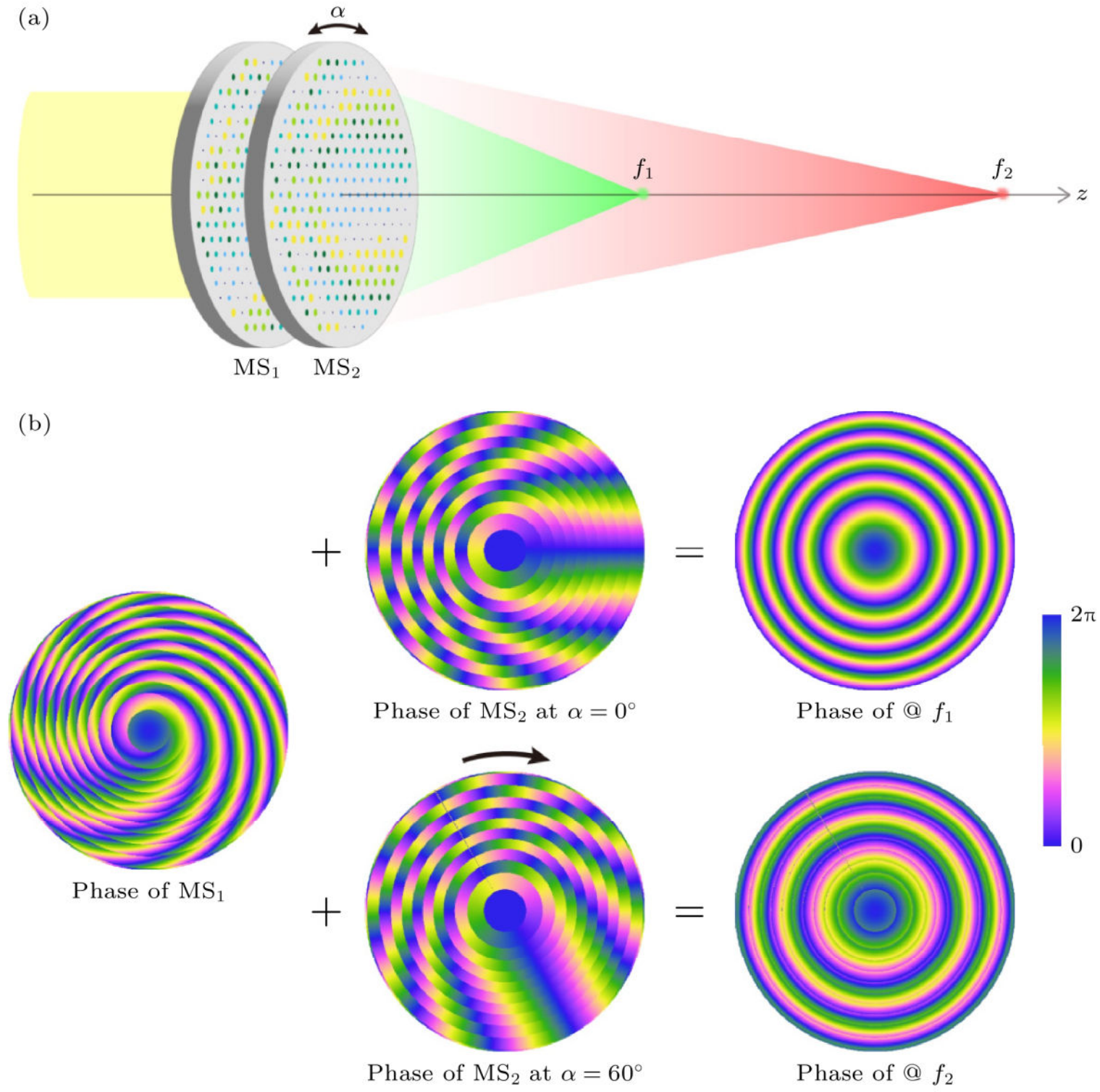

**Fig. 1 Schematic of the zooming principle of a bilayer metalens based on rotation-controlled phase: (a) Schematic of the bilayer rotational zoom metalens. The focal length can be tuned by rotating the second metasurface; (b) Illustration of the phase distributions of the two metasurfaces and their combined phase. As the rotation angle increases, the number of equiphase rings decreases, corresponding to an increase in the lens focal length.**

# 3 Design and Simulation of Metasurface Unit Structures

To realize the target phase distribution required for the dual-layer rotational zoom metalens, it is necessary to construct metasurface unit structures that exhibit both high transmittance and a complete phase modulation range within the operating frequency band. This study selects alumina ceramic with a refractive index of $n \approx 3$ as the dielectric material and employs cylindrical dielectric pillars as the basic structural units of the metasurface, as illustrated in Fig. 2(a). The cylindrical dielectric units possess rotational symmetry, rendering their electromagnetic response independent of the polarization direction of the incident electromagnetic wave. Consequently, the metalens designed in this study is polarization-independent. To balance the phase modulation range with transmission efficiency, we performed parameter scans on cylindrical arrays arranged in a triangular lattice at an operating frequency of 10 GHz. We examined the effects of the lattice constant $a$, cylinder height $h$, and diameter $D$. Figures 2(b) and 2(c) demonstrate the impact of different lattice constants $a$ and cylinder diameters $D$ on transmittance and phase response under a fixed cylinder height $h$. Conversely, Figs. 2(e) and 2(f) present the scanning results for cylinder height $h$ and diameter $D$ with a fixed lattice constant $a$. The results indicate that the cylinder diameter $D$ is the primary parameter determining phase modulation capability. As the diameter increases, the phase response rises monotonically, achieving continuous coverage of $0$—$2\pi$. However, excessive structural height introduces Fabry-Perot type resonances, leading to reduced transmittance. Considering both transmission efficiency and phase coverage, we selected a unit structure combination with $a$ = 18 mm and $h$ = 24 mm as the design scheme. This configuration ensures transmittance above 85% and continuous $2\pi$ phase response coverage. The relationship between transmittance, phase response, and cylinder diameter is shown in Fig. 2(d). This relationship establishes the phase library required for subsequent lens design. Based on this phase library, we can precisely map the design phase distribution of the dual-layer metasurface to specific geometric parameters, thereby realizing the overall structural design of the dual-layer rotational zoom metalens.

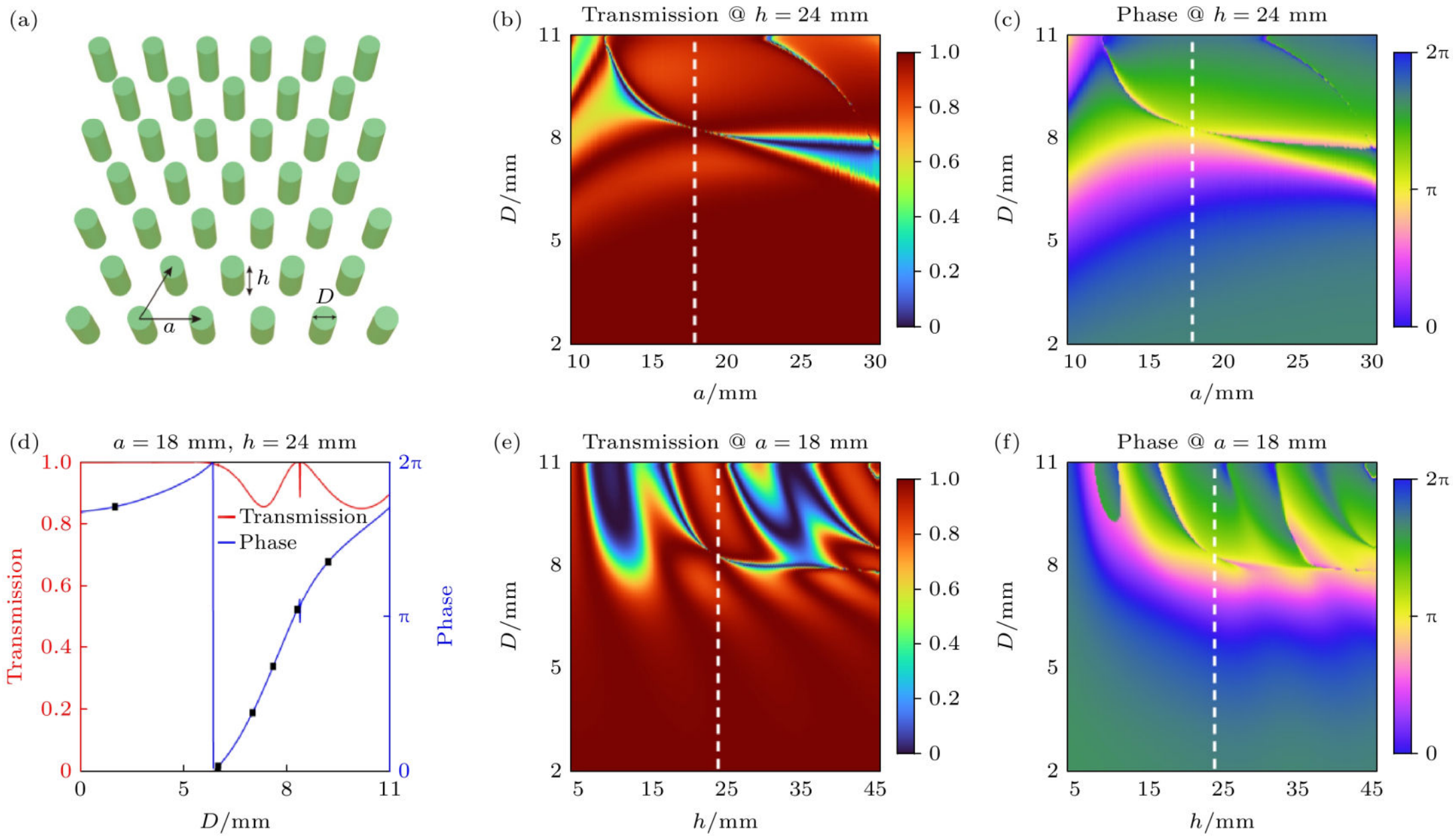

**Fig. 2 Unit design and parameter optimization of the metalens: (a) Schematic of the unit cell of metasurface and its geometric parameters; (b), (c) Transmission and phase responses under different lattice constants and cylinder diameters at a height of *h* = 24 mm; (d) Transmission and phase distributions for the selected parameters, with black dots indicating the final chosen unit cell parameters; (e), (f) Transmission and phase responses for different cylinder heights and diameters at a lattice constant *a* = 18 mm.**

To validate the theoretical model and design methodology, we performed full-wave simulations of the proposed bilayer rotational zoom metalens using the finite-difference time-domain method. Figure 3(a) presents the electric field intensity distribution in the depth-of-focus plane (*XZ* plane) under various rotation angles, visually demonstrating the shift in focal position with changing rotation angle. To compare the simulation results with theoretical expectations, we matched the ideal lens phase function (Equation (8)) with the combined phase profile. This process yielded the theoretical curve for focal length variation with rotation angle, shown as the black dashed line in Figure 3(a). The simulation results align with theoretical predictions, confirming the zoom characteristics of the design. Figure 3(b) displays the electric field intensity distribution in the focal plane (*XY* plane). The results show concentrated focal energy, a clear main peak, and a symmetric distribution, indicating that the metalens maintains excellent focusing performance across different rotation angles. Simulation results indicate that as the rotation angle $\alpha$ increases from -60° to 60°, the focal length of the lens varies from 99 mm to 298 mm. This corresponds to a zoom range of approximately threefold, while the numerical aperture decreases from 0.83 to 0.45. These results are consistent with the theoretical model expectations, fully verifying the feasibility and accuracy of the proposed design method for the large *NA* bilayer rotational zoom metalens.

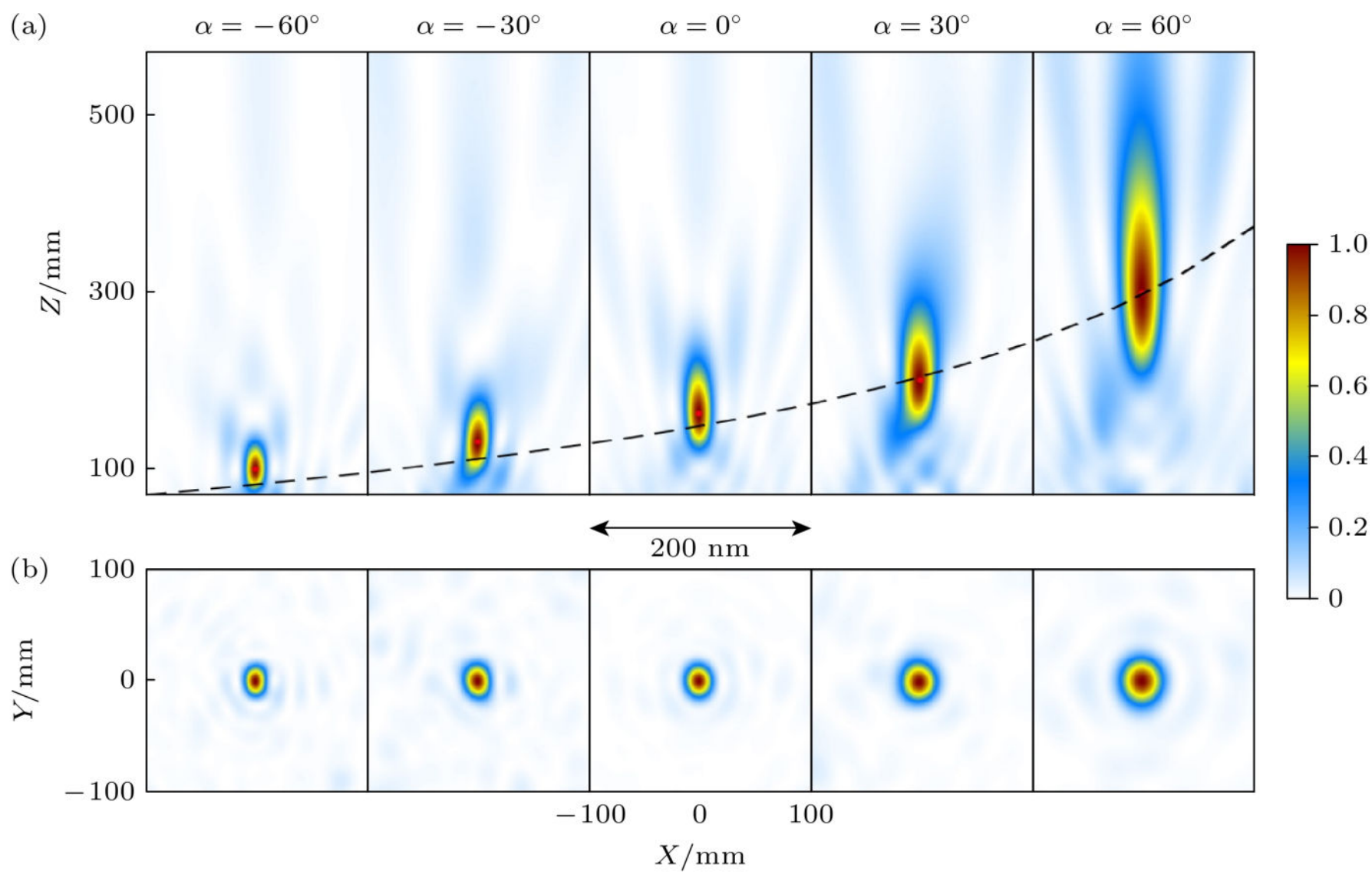

**Fig. 3 Full-wave simulation results of the bilayer rotational zoom metalens: (a) *XZ*-plane field intensity distributions at different rotation angles. The dashed line represents the theoretically calculated focal length variation; (b) Field intensity distribution on the focal plane (*XY* plane), showing the spatial distribution and variation trend of the focal spot energy. The system exhibits stable focusing characteristics.**

# 4 Experimental Results

To further verify the actual performance of the dual-layer rotating zoom metalens, we fabricated a sample using alumina ceramic pillars based on the aforementioned phase library design. Figure 4(a) displays six dielectric pillar unit cells with different diameters (3-9.2 mm) used to construct the metalens. The cylindrical height is $h$ = 24 mm. These unit cell parameters correspond one-to-one with the phase library shown in Figure 2(d). Figures 4(b) and 4(c) present photographs of the dual-layer metalens. The lens aperture is 300 mm. The ceramic pillars are embedded in a foam substrate with a refractive index of $n \approx 1.05$. They are arranged in a triangular lattice with a lattice constant of 18 mm. The operating frequency is 10 GHz, corresponding to a wavelength of 30 mm. To facilitate the identification of different dielectric pillar unit cells, the photographs use the color codes shown in Figure 4(a) to represent different cylinder diameters. Figure 4(d) illustrates the schematic of the experimental system. In the experiment, a low-sidelobe horn antenna serves as the excitation source to generate an approximate plane wave incident on the metalens surface. The two metalens layers are aligned, closely attached, and fixed on a bracket. Their central plane is defined as $z = 0$. An electric field probe is mounted at the end of a three-dimensional precision mechanical arm. It can move with high precision along the $x$, $y$, and $z$ directions. This setup enables point-by-point scanning of the spatial electric field to acquire three-dimensional electric field distribution information.

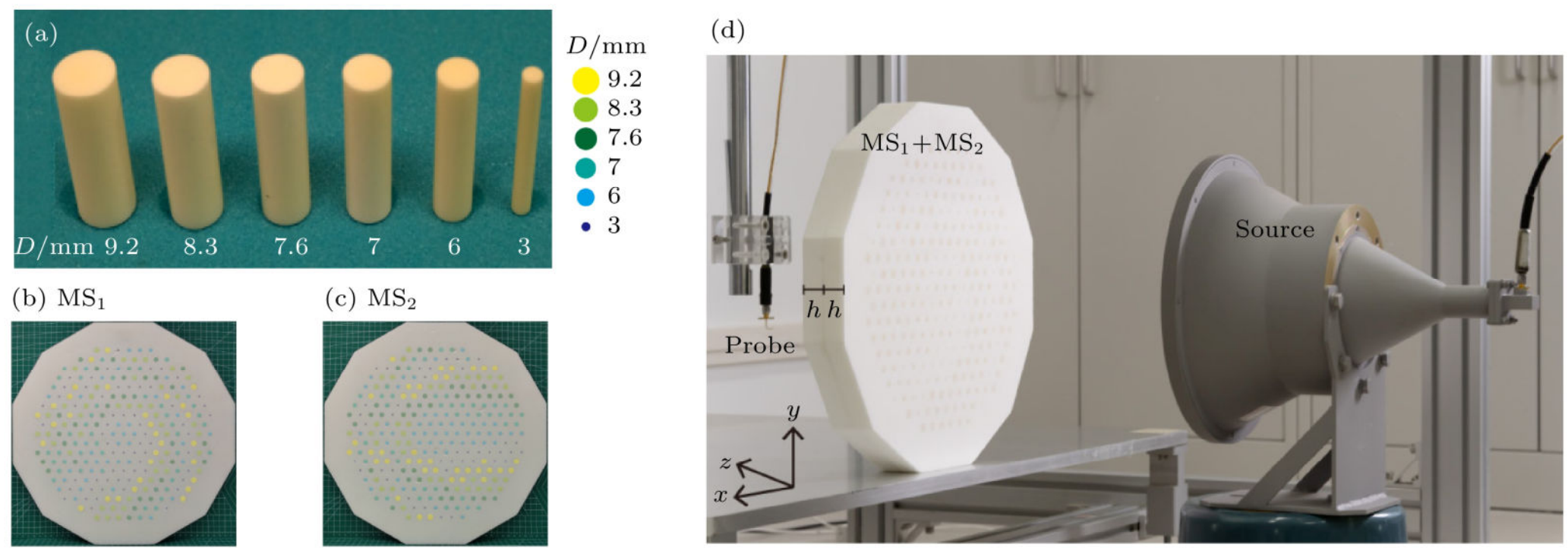

**Fig. 4 Experimental setup and sample of the bilayer rotational zoom metalens: (a) Six types of ceramic cylinder units corresponding to the phase library; (b), (c) Photographs of the bilayer metalens (aperture 300 mm); (d) Photo of the near-field electromagnetic scanning experimental system, with the bilayer metalens closely attached.**

Figure 5(a) presents the experimental measurements of the depth of focus (*XZ* plane) under various rotation angles. As the rotation angle $\alpha$ of the second metasurface layer $MS_2$ increases, the focal point gradually moves away from the lens surface. This clearly demonstrates the tunable nature of the focal length. Figure 5(b) shows the experimental results for the focal plane (*XY* plane). The energy is concentrated in the focal region with a distinct main peak, indicating that the sample maintains stable focusing capability during actual measurements. To further analyze the variation in focal length and focusing performance, we compare the experimental results with simulation data. As shown in Figure 5(c), the trend of the experimentally measured focal length versus rotation angle aligns with the theoretical prediction curve. Although there are slight deviations in the specific focal length values, these discrepancies primarily stem from non-ideal factors. These include deviations in the geometric dimensions and refractive index of the sample, as well as differences between the incident electromagnetic waves in the experiment and ideal plane wave conditions. Within the range of −60° to 60°, the focal length increases from approximately 92.5 mm to 322.5 mm, achieving a zoom range of about three times. Furthermore, as illustrated in Figure 5(d), the variation trend of the full width at half maximum (FWHM) of the focal spot with respect to the rotation angle agrees well with the simulation results. This further confirms the high applicability of the model under large *NA* conditions.

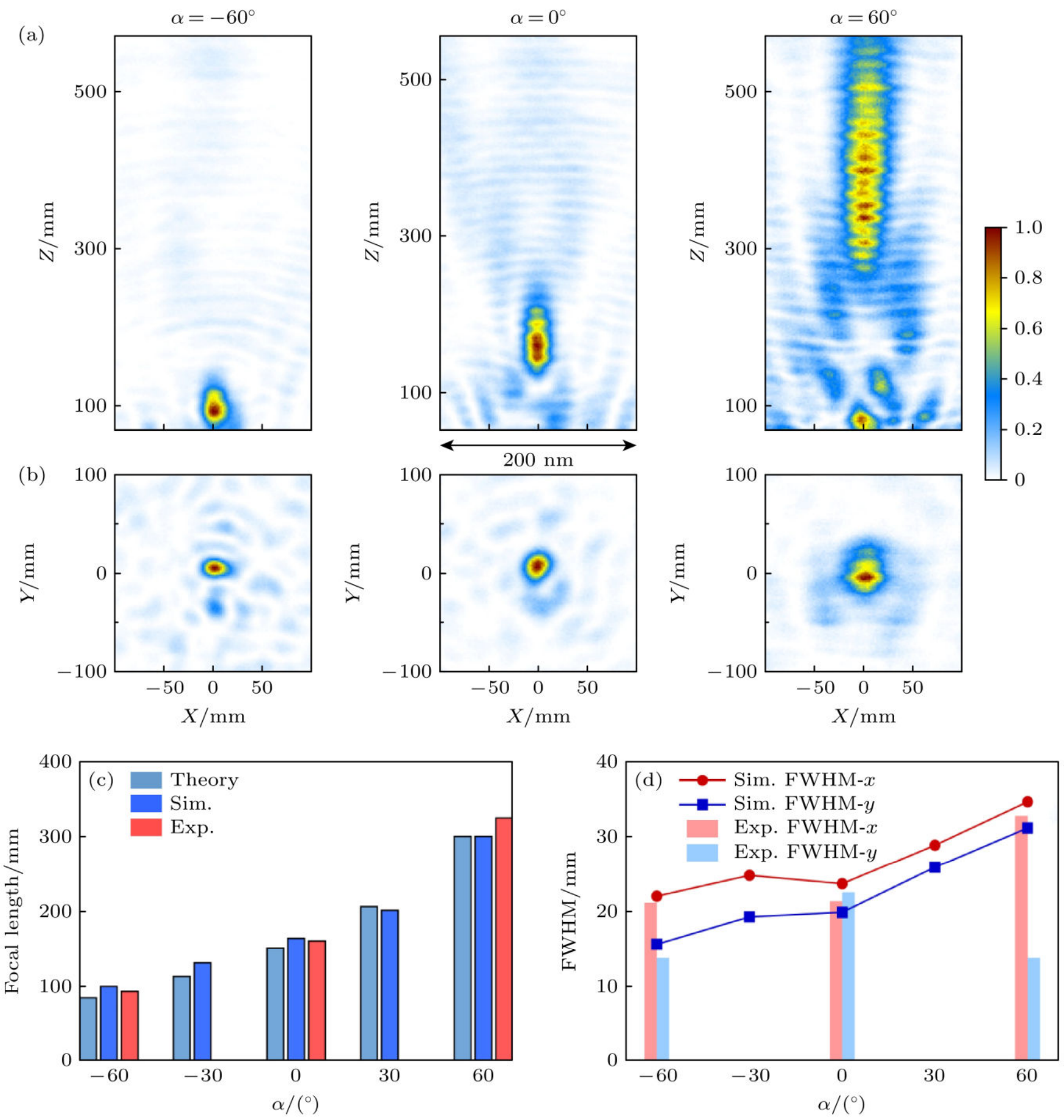

**Fig. 5 Experimental results of the bilayer rotational zoom metalens: (a) Focal depth plane (*XZ*-plane) field intensity distributions at different rotation angles; (b) Experimental results on the focal plane (*XY*-plane); (c) Comparison of focal length variation trends among theory, simulation, and experiment; (d) Comparison of simulated and measured focal spot FWHM.**

## 5 Conclusion

To address the application requirements of traditional bilayer Moiré rotational zoom metalenses under large numerical aperture conditions, this paper proposes a design method for a rotatable zoom bilayer metalens with a large numerical aperture. Unlike traditional models that rely on parabolic lens phase approximations, our approach starts from the ideal aberration-free lens phase function. This method theoretically reduces phase deviation in non-paraxial regions under large numerical aperture conditions. It helps suppress aberrations caused by high-order phase errors, enabling the rotational zoom system to maintain stable focusing performance over a wide angular range. Based on this, we completed the structural design and numerical simulation of the bilayer rotational zoom metalens. We systematically verified the feasibility of the proposed design method through finite-difference time-domain simulations and near-field scanning experiments. The results indicate that at an operating frequency of 10 GHz, the focal length varies within the range of 100—300 mm as the rotation angle of the second metasurface increases from −60° to 60°. The zoom ratio is approximately 3:1, and the trend of focal length variation aligns with theoretical analysis and numerical simulation results. It is important to note that due to the strong dispersion of the phase response of the metasurface units, the focal spot of this bilayer metalens deteriorates significantly at off-center operating frequencies (such as 9 GHz or 11 GHz). Therefore, achromatic design schemes are required in future work to optimize the frequency bandwidth. Throughout the zooming process, the focal shape remains stable with concentrated energy distribution, verifying the effectiveness of the proposed phase model and design method. In summary, this study establishes a complete research framework ranging from ideal phase theory derivation and bilayer metasurface unit design to experimental verification. We achieved mechanical-displacement-free rotational zooming with a large numerical aperture, reaching a maximum numerical aperture of 0.83. This research provides new theoretical references for designing rotational zoom Moiré metalenses under large *NA* conditions. It also offers valuable insights for the development of lightweight, planar, and tunable focusing optical systems.